\documentclass[runningheads,a4paper]{llncs}

\usepackage{amssymb}
\usepackage{graphicx}

\usepackage{url}
\usepackage{cite}
\urldef{\mailsa}\path|{elena.stoll2, kammer}@htw-dresden.de|    
\newcommand{\keywords}[1]{\par\addvspace\baselineskip
\noindent\keywordname\enspace\ignorespaces#1}

\begin{document}

\mainmatter  

\title{Online Knowledge Base for Designing Shape-changing Interfaces using Modular Workshop Elements\let\thefootnote\relax\footnotetext{This paper has been peer-reviewed and accepted to the \textit{Workshop on Visual Interface Design Methods (VIDEM 2020)}, held in conjunction with the International Conference on Advanced Visual Interfaces (AVI 2020). Ischia, Italy. Workshop organizers: Mandy Keck, Dietrich Kammer, Alfredo Ferreira, Andrea Giachetti, Rainer Groh, http://videm.mediadesign-tud.de/}}

\titlerunning{Online Knowledge Base for Designing Shape-changing Interfaces}

%
%
\author{Elena Stoll \and Dietrich Kammer}
\authorrunning{Stoll \& Kammer}

\institute{University of Applied Sciences Dresden\\
Chair for Visual Engineering\\
Friedrich-List-Platz 1, 01069 Germany\\
\mailsa\\
\url{https://htw-dresden.de}}

%
%

\toctitle{Lecture Notes in Computer Science}
\tocauthor{Authors' Instructions}
\maketitle

\begin{abstract}
Building and maintaining knowledge about specific interface technologies is a challenge. Current solutions include standard file-based document repositories, wikis, and other online tools. However, these solutions are often only available in intranets, become outdated and do not support the acquisition of knowledge in an efficient manner. The effort to gain an overview and detailed knowledge about novel interface technologies can be overwhelming and requires to research and read many technical reports and scientific publications. We propose to implement open source online knowledge bases that include building blocks for creating custom workshops to understand and apply the contained knowledge. We demonstrate this concept with a knowledge base for shape-changing interfaces (SCI-KB). The SCI-KB is hosted online at GitHub and fosters collaborative creation of knowledge elements accompanied by practical exercises and workshop elements that can be combined and adapted by individuals or groups of people new to the topic of shape-changing interfaces.

\keywords{Shape-changing interfaces, Knowledge Base, Design Methods, Workshops}
\end{abstract}

\section{Introduction}
Keeping knowledge resources up-to-date and accessible is a particular challenge in areas that are in a constant state of change, such as specific novel interface technologies. Moreover, if knowledge is not preserved properly or gets lost, e.g. due to staff changes, the effort to gather information increases significantly. Re-acquiring knowledge or familiarizing yourself with a new topic is often inefficient. Not only because it takes considerable time, but in practice this usually means reading a large number of scientific or technical documents, as there are scarcely tutorials or exercises to apply the contained knowledge. A wide range of standard file-based document repositories, wikis, and other online tools is already in use to aggregate company and research knowledge. However, these solutions are often only available in intranets, become outdated, or do not support the acquisition of knowledge in an efficient manner. Even worse, some collections are still maintained in private documents or folders by individuals. We therefore propose a modular structure for open source online knowledge bases containing building blocks to create custom workshops. These workshops can serve to acquire and apply knowledge through creativity techniques and design methods. Our concept is demonstrated with a knowledge base for shape-changing interfaces (SCI-KB), which is hosted online at GitHub\footnote{SCI-KB website https://visualengineers.github.io/sci-knowledge-base/, Last access: 17.09.2020}. The SCI-KB is designed to be an effective tool for interdisciplinary teams to explore the design space of shape-changing interfaces. It fosters collaborative creation of knowledge elements coupled with practical exercises and workshop elements that can be arranged and modified by individuals or groups of people. 

\section{Background and Related Work}

The motivation for this work on an online knowledge base is the extensive recent research in the areas of tangible interaction \cite{HCI-026} and organic user interfaces \cite{10.1145/1349026.1349033}. For instance, human hands are universal tools with a potential that is not yet sufficiently exploited by human-computer interaction (cp. \cite{Kammer2018iCom}). The research areas mentioned above play a crucial role in fully exploiting this potential. Interaction becomes more direct and the use of understandable metaphors from reality is possible \cite{10.1145/1124772.1124965, 10.1145/1357054.1357089}. The overall goal is to achieve information transfer by providing rich feedback in the form of tangible changes of the interaction elements, mitigating the need for visual perception. Thus, interaction becomes a multi-sensory experience that takes place in spatial dimensions (cf. \cite{10.1145/2207676.2207781}). Shape-changing interfaces (SCI), which can change their physical form, play a decisive role in this context. Although these technologies promise an appealing and diverse user experience, they also lead to new challenges in interaction design. For example, designing ever-changing, volatile interfaces poses the question, which aspects can be designed at all. The third dimension allows users to immerse themselves in their data spaces without the use of virtual reality technologies, which requires the investigation of feasible haptic encodings, metaphors, and application scenarios. Elastic displays, as a sub-type of shape-changing interfaces, essentially consist of a deformable surface that automatically returns into its initial flat state after interaction. The surface is often made of fabric, but also of gel \cite{10.1145/2254556.2254719} or other materials. In contrast to conventional multi-touch surfaces, interaction occurs through manual deformation registered by sensors, for example depth cameras. In current hardware solutions, digital content is usually displayed using video projectors. In summary, on an elastic display, users can carry out true three-dimensional inputs by pulling or pushing the surface, which becomes a haptic experience due to the force-feedback of the fabric. 
Recently, scientists requested the consolidation of knowledge about shape-changing interfaces \cite{10.1145/3173574.3173873, 10.1145/2207676.2207781}. As the interaction with such interfaces is based on arbitrary input via human hands, a lot of knowledge is implicitly created by designers and developers. The distinction between implicit (subjective) and explicit (objective) aspects of knowledge is derived from Polanyi\cite{Hedesstrom2000WhatIM}. Polanyi states that despite theoretical and objective aspects, every knowledge contains a certain non-explicable, personal component concerning practical and bodily skills. Existing knowledge bases are mainly designed to contain explicit, factual knowledge using documents. Recent research that integrates more subjective knowledge into knowledge bases focuses on approaches such as subjective enrichment of knowledge via crowd-sourcing, semantic search, Q\&A systems, or natural language translation \cite{10.1145/3183713.3183732}. The SECI model offers a theoretical solution to combine both types of knowledge by considering knowledge growth to be the transformation of knowledge states\cite{10.2307/41165942}. However, this model has to be adapted to specific use cases of knowledge management to be of productive use.

\section{Shape-changing Interfaces Knowledge Base}

The Shape-changing Interfaces Knowledge Base is part of an ongoing research project (ZELASTO), which explores the interaction with complex data using zoomable user interfaces on elastic displays. Although our focus is currently on elastic displays only, we hope to expand the knowledge base in the future for other forms of shape-changing interfaces and hardware.

Compiling and persisting the knowledge created by the team members was a challenge for several reasons. Depending on their prior experiences, each team member needed individual information to achieve a common level of knowledge at the beginning of the project. Knowledge was scattered within the team in any manner imaginable and experienced colleagues were sometimes unaware that their accumulated personal knowledge from research practice is not self-evident and therefore of great value to others. Moreover, the interdisciplinary nature and novelty of the technology requires specific interface and interaction solutions, which are not readily available today. We found that currently there is a lack of resources that link the various disciplines with each other, efficiently tackle practical issues, and support the dissemination of subjective knowledge.

For instance, the team works at different locations at different times. For this purpose, a central knowledge pool had to be created that was always accessible from remote locations. Considering the Covid-19 pandemic of 2020, our approach has proven to be more than crucial for continued work under unforeseen circumstances and restrictions. To avoid burdening the workflow with additional documentation, the solution had to be seamlessly integrated into the existing style of work. 

With the SCI-KB, we explore novel paths of collaboration by efficiently and effectively increasing the level of both implicit and explicit knowledge on the design of shape-changing interfaces for every team member. Therefore, the architecture for storing knowledge items on shape-changing interfaces has to be highly flexible regarding access, expansion, and management.

\begin{figure}
\centering
\includegraphics[width=\columnwidth]{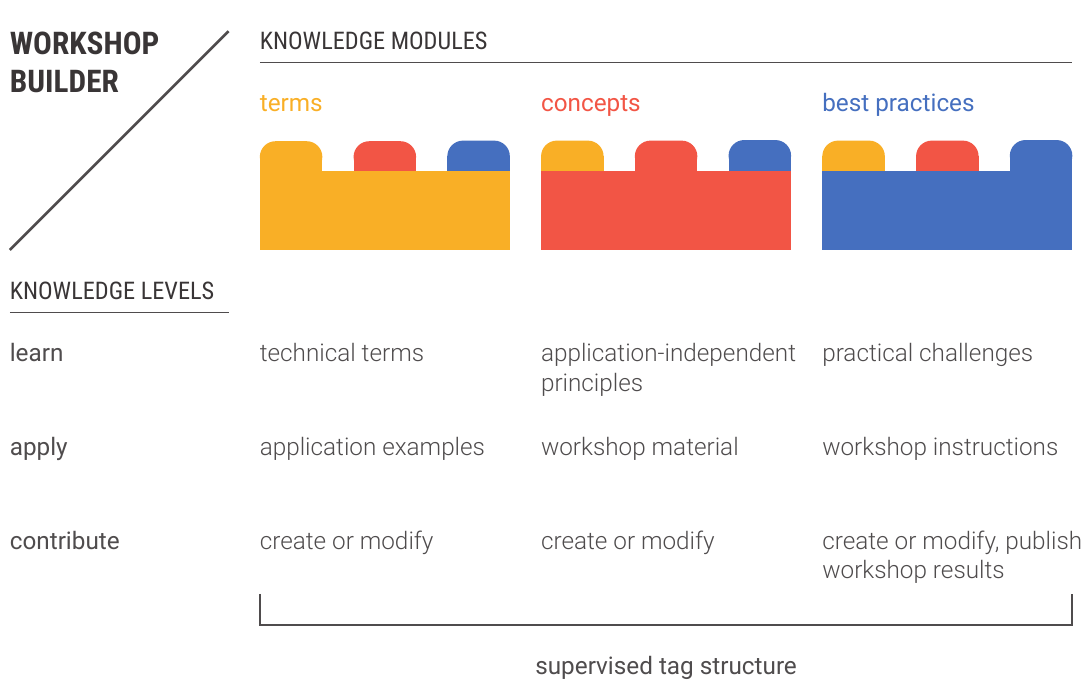}
\caption{Knowledge items are categorized using three knowledge modules (terms, concepts, best practices) in the context of three knowledge levels (learn, apply, contribute). Knowledge items can then be combined across these modules and levels to create individual workshops.}
\label{fig:sci-kb}
\end{figure}

\subsection{Concept} \label{concept}
The SCI-KB addresses personal backgrounds and knowledge levels with regards to the topic of shape-changing interfaces. Hence, knowledge items are stored in three distinct modules. Arbitrary connections of items from different modules are possible, creating an interwoven three-level-architecture, which is the core of the SCI-KB (see Fig. \ref{fig:sci-kb}). 

The knowledge modules \textit{Terms}, \textit{Concepts}, and \textit{Best Practices} can be arranged in any combination and pervade each of the three knowledge levels \textit{Learn}, \textit{Apply}, and \textit{Contribute}. The modules and levels of knowledge are inspired by Bloom's taxonomy of educational objectives \cite{10.1207s15430421tip4104} as well as Nonaka's SECI model for knowledge construction\cite{10.2307/41165942}. Similarly, we also address factual knowledge on the lower levels (\textit{Learn}, \textit{Terms}, \textit{Concepts}) and procedural knowledge on the higher levels (\textit{Contribute}, \textit{Apply}, \textit{Best Practices}), while emphasizing the interplay of practice and theory for both individuals and groups.

The core is supplemented by additional resources such as paper references, links, and interviews and is thematically structured using tags. To provide an overview of the knowledge base, the content is framed by a landing page, project news page, and a presentation of contributors. The \textit{Workshop Builder} provides assistance in finding an arrangement of knowledge items, using them as building blocks to conduct a workshop that is tailored to individual needs. The links between knowledge items create a knowledge network from which building blocks for workshops can be extracted as needed.

In the following, we explain the three modules for the knowledge items as well as the supervised tag structure and the workshop builder.

\subsubsection{Knowledge Module Terms.}
On the \textit{Learn} level, each knowledge item explains a technical term for shape-changing interfaces, for example ''Elastic Display``. On the \textit{Apply} level, examples show how the term is used in practice. The \textit{Contribute} level invites users to modify or add terms as knowledge items.

\subsubsection{Knowledge Module Concepts.} 
In the style of a cheat sheet, each concept on the \textit{Learn} level explains an application-independent principle that can be useful for the implementation of a shape-changing interface. An inspiration for this approach are the visualization cheat sheets by Wang et al. \cite{10.1145/3313831.3376271}. A concept includes, for example, design guidelines, overviews of chart and data types, or a gesture catalogue for interaction with elastic displays. On the \textit{Apply} level, concepts provide workshop material. Just as a term, a concept can be modified or created on the \textit{Contribute} level.

\subsubsection{Knowledge Module Best Practices.} 
On the \textit{Learn} level, a best practice shows how to overcome a practical challenge that may arise in the design process of shape-changing interfaces, such as the choice of a suitable prototyping technology. On the \textit{Apply} level, we provide workshop instructions using materials particular to the concept. Participants are invited to publish the results of their workshops on the \textit{Contribute} level or to create their own best practice.

\subsubsection{Tags.} 
With the help of the supervised tag structure, knowledge items are grouped in the four areas proposed by Alexander et al. for the classification of the research field concerning shape-changing interfaces (technological, behavioural, design and societal) \cite{10.1145/3173574.3173873}. This division both separates disciplines within the field and reflects successive stages of implementation. In this way, a thematic access to the knowledge base is possible. The tag structure is supervised by the maintainers of the knowledge base to ensure its consistency and conciseness.

\subsubsection{Workshop Builder.} 
With the Workshop Builder, modular workshops tailored to custom needs can be composed from knowledge items. Based on the context of the workshop and its objective, the Workshop Builder suggests one of the following starting points for a practical acquisition of knowledge: overview of topics (thematic), keyword search along with \textit{Terms} (deep-dive), \textit{Concepts} (method-based), \textit{Best Practices} (problem-oriented), or an overview of knowledge modules (explorative). Additionally, formats such as interviews can provide users with practical advice during the planning phase of a workshop.

\subsection{Implementation}

The SCI-KB is an open-source project on the collaborative software development platform GitHub. Emphasizing its open nature and flexible structure, the website can be accessed and edited online at any time with each page providing editing links to GitHub. Changes are integrated by the maintainers of the repository, following GitHub's lightweight, branch-based workflow. GitHub Pages allows the rapid publication of GitHub-hosted projects as a website, using markdown files and customizable templates. The SCI-KB is based on the Docsy Template\footnote{Docsy Template https://github.com/vsoch/docsy-jekyll, Last access: 15.04.2020}, which is especially suited for online documentation. In addition, we adapted stylesheets and the structure of the template and we created custom templates with reusable code fragments to simplify the publication of knowledge items.

Content is continuously integrated into the SCI-KB as new knowledge items to document and disseminate knowledge. Initially, 48 knowledge items were created (26 terms, 11 concepts and 11 best practices). Some of them are not completely finalized, which reflects the status of our ongoing research. For example, workshop instructions for best practices can only be derived as soon as new findings are available. 

Basic functionalities of the Workshop Builder are implemented. Users receive recommendations for knowledge modules according to their needs by answering multiple choice questions about their goals, providing the number of participants as well as their working locations, and selecting one out of five statements that best reflects their situation. Based on this information, an entry point for one of the walkthroughs described in section \ref{concept} is recommended. A supplementary interview covers practical issues in conducting workshops. In the future, it is planned that a schedule for the workshop can be created by stacking the knowledge items (see section \ref{conclusions}).

\begin{figure}
\centering
\includegraphics[width=\columnwidth]{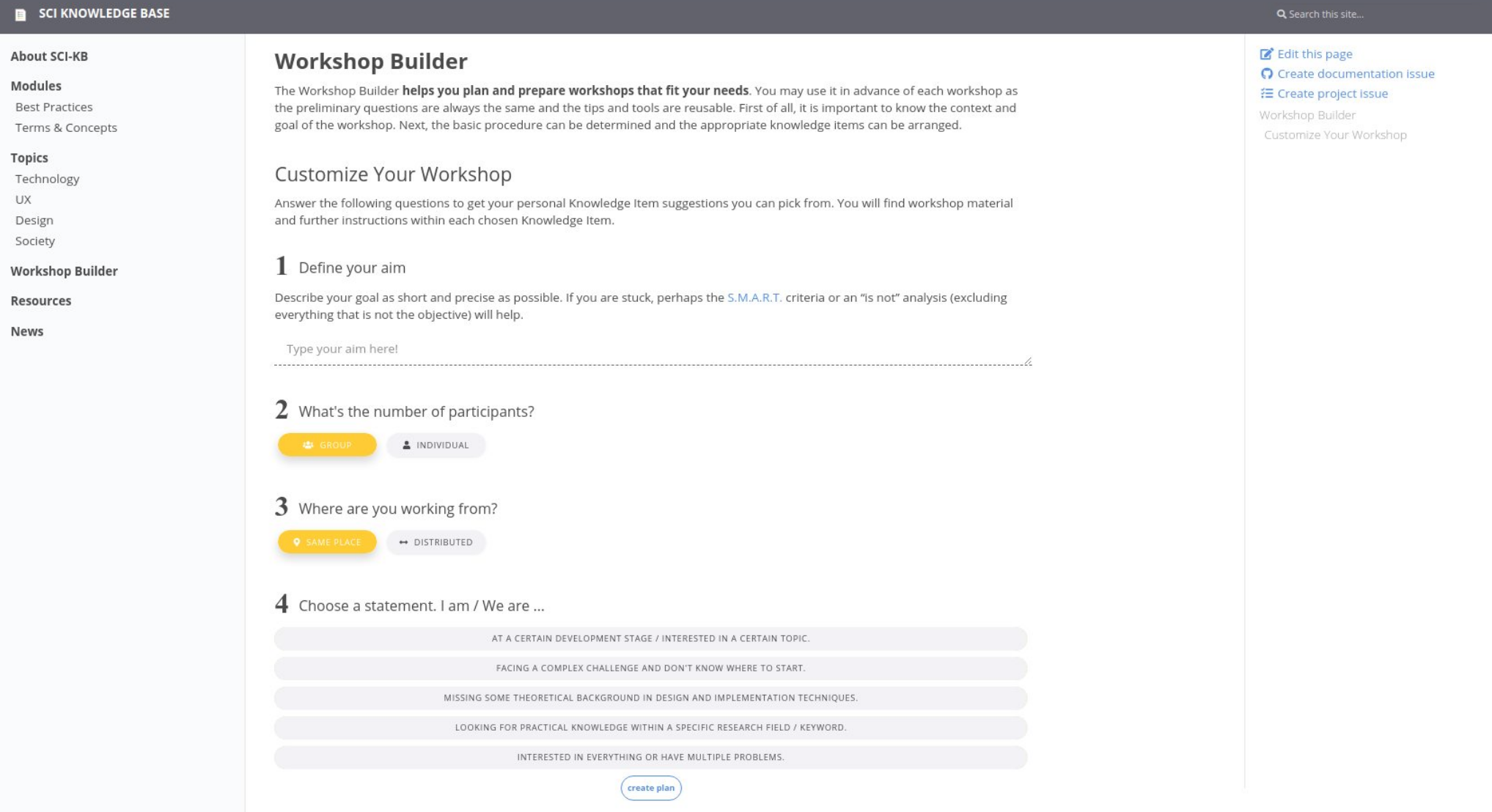}
\caption{Screenshot of the workshop builder.}
\label{fig:workshop-builder}
\end{figure}

\subsection{Demonstration}

In the following, the flexibility of the modular usage concept is demonstrated for two workshop planning scenarios. The scenarios illustrate how information on the workshop context provided by the user (Input) leads to the Workshop Builder's suggestions for individual content (Output). Three aspects are taken into account to calculate the output: (1) the number of participants, (2) the location of participants and (3) a statement about the workshop context. Finally, we clarify how the output contributes to the selection of modules and tools from which individual workshops can be composed.

\subsubsection{Example 1.}
A group of designers, programmers, and architects would like to evaluate to what extent shape-changing interfaces can support proper public communication for construction projects. Preferably, a face-to-face meeting is to be avoided, as the participants work in different places. 
\begin{itemize}
  \item Input. (1) Group, (2) Distributed, (3) "We are facing a complex challenge and don't know where to start".
  \item Output. (1,2) List of collaboration tools for online workshops, (3) Problem-oriented walkthrough starting at best practices.
\end{itemize}

The collection of best practices contains two that fit the use case. To overcome the challenges for the specific use case, the examples and workshop instructions may be adapted and updated. Related technical terms and workshop methods are linked to efficiently provide factual knowledge.

\subsubsection{Example 2.}
A user experience designer, who is inexperienced with elastic displays is hired to improve the user experience (UX) of an elastic display.

\begin{itemize}
  \item Input. (1) Individual, (2) Same Place, (3) "I am interested in a certain topic".
  \item Output. (1,2) List of tools to consolidate practical skills on your own, (3) Thematic walkthrough starting with an overview of topics.
\end{itemize}

All knowledge items related to aspects of user experience are contained within this topic. A range of examples supports a fast, broad knowledge acquisition. The UX designer may also create new items, for example a \textit{Concept}, to describe the applicability of a common UX method for the development of applications on an elastic display.

\section{Conclusions and Future Work}
\label{conclusions}

In this paper, we tackled the challenge of creating a modular, open-source knowledge base suited for academia, industry, and interdisciplinary collaboration. Our current focus is on shape-changing interfaces, particularly elastic displays. With this knowledge base, we hope to instigate collaboration across research groups and companies on this topic. However, we also aim to create more knowledge bases for different topics based on current and future research interests.

With the first prototype of our Workshop Builder, we have shown how a passive accumulation of knowledge can be turned into a platform for active learning. The dissemination of subjective knowledge can be facilitated further by expanding the space for personal experiences, for example through integrating more workshop content, tools the contributors find useful, or alternative formats such as interviews. Hence, workshop material for 3D printing or for prototyping with paper or platforms such as Arduino should be provided. Similar to recommendation systems, we plan to suggest individual workshop plans based on selected knowledge items.

Another valuable addition to the SCI-KB is the integration of software artefacts. For instance, an emulator, which simulates events from an elastic display, could be used for prototyping in a workshop. Linking and referencing repositories on GitHub with entire libraries that help programmers in implementing their own applications for elastic displays is also conceivable.

Finally, the overall access to knowledge can be improved, e.g. by providing teaser texts and hierarchical or aggregated information landscapes to gain a quick overview of the content.

\section*{Acknowledgements}
This work has been supported by the European Regional Development Fund and the Free State of Saxony (project no. 100376687). Thanks are due to the ZELASTO project team members: Lukas B\"uschel, Stefan Gehrke, Mathias M\"uller, Martin Herrmann, Josefine Zeipelt, and Franziska Hannß. Daniel Redetzki also contributed contents to the knowledge base.

\bibliographystyle{splncs04}
\bibliography{mybibliography}

\end{document}